\documentclass[onecolumn,preprint,showpacs,superscriptaddress]{revtex4-1}
\usepackage{amssymb}
\usepackage{amsfonts}
\usepackage{amsmath}
\usepackage{graphicx}
\usepackage{epstopdf}
\usepackage{float}
\usepackage{relsize}

\begin{document}

\title{ Quantum dynamics of scalar particles in the space-time of a cosmic string in the context of gravity's rainbow}
\author{L. C. N. Santos}
\email{luis.santos@ufsc.br}
\affiliation{ Departamento de F\'isica, CCEN - Universidade Federal da Para\'iba; C.P. 5008, CEP  58.051-970, Jo\~ao Pessoa, PB, Brasil}
\author{C. E. Mota}
\email{clesio200915@hotmail.com}
\affiliation{Departamento de F\'{\i}sica - CFM - Universidade Federal de Santa Catarina, CP. 476
- CEP 88.040 - 900, Florian\'{o}polis - SC - Brazil}
\author{C. C. Barros Jr.}
\email{barros.celso@ufsc.br }
\affiliation{Departamento de F\'{\i}sica - CFM - Universidade Federal de Santa Catarina, CP. 476
- CEP 88.040 - 900, Florian\'{o}polis - SC - Brazil}
\author{L. B. Castro}
\email{lrb.castro@ufma.br}
\affiliation{Departamento de F\'{\i}sica, Universidade Federal do Maranh\~ao (UFMA), Campus Universit\'ario do Bacanga, 65080-805, S\~ao Lu\'{\i}s, MA, Brazil}
\affiliation{Departamento de F\'{\i}sica e Qu\'{\i}mica, Universidade Estadual Paulista (UNESP), Campus de Guaratin\-gue\-t\'{a}, 12516-410, Guaratinguet\'{a}, SP, Brazil}
\author{V. B. Bezerra}
\email{valdirbarbosa.bezerra@gmail.com}
\affiliation{ Departamento de F\'isica, CCEN - Universidade Federal da Para\'iba; C.P. 5008, CEP  58.051-970, Jo\~ao Pessoa, PB, Brasil}
\begin{abstract}

We analyze the dynamics of scalar particles in gravity's rainbow considering the space-time of a cosmic string in this modified gravity. Thus, we solve the Klein-Gordon equation for two types of potential in which two possible rainbow functions are selected. In all cases studied we obtain exact solutions for the wave equation, derive the energy spectrum, analyze it in detail and discuss the dependence of the results with the choice of the rainbow functions.

\end{abstract}

\maketitle

\preprint{}

\volumeyear{} \volumenumber{} \issuenumber{} \eid{identifier} \startpage{1} 
\endpage{10}

\section{Introduction}
\label{sec1}

Recent observations of gravitational waves predicted by Einstein \cite{waves1} and shadows of black holes \cite{black1}  reinforce the idea that General Relativity (GR) is a fundamental theory for the description of gravitational phenomena. In principle, any gravitational theory which purports to describe the nature of space-time should agree with GR in regimes where it is well tested. Observations of supernova explosions in 1998 suggest that the expansion of the universe is accelerating rather than slowing down as some models have predicted. As a consequence of this observation, on the large scale regime, GR may be maintained as the standard theory of the space-time if one consider the presence of some unknown constituent as dark energy, which should be responsible for this acceleration and dark matter. Another approach to dealing with the open problems in cosmology and astrophysics is to consider modifications \cite{fder1,santos8} in the usual GR theory that preserve the well-tested results. In recent years, there has been a growing awareness that modifications in GR  may provide satisfactory responses to the cosmological observations that reveal the existence of dark energy. Certain modified gravity theories may provide a plausible explanation for the problems of dark matter and dark energy through additional degrees of freedom. 

Since GR has been formulated it was discovered that the geometry of space-time may be associated with physical quantities through the energy-momentum tensor. We also know that factors associated with topology of the space-time can also influence the dynamics of matter or light. As an example, we can mention the space-time around a cosmic string \cite{string2,string3} which is characterized by an angular deficit that is determined by the linear mass density of the cosmic string. This line of the energy-momentum tensor is termed a topological defect. In this space-time, which is locally flat but not globally flat, particles can be deflected even considering that the gravitational potential of a straight string vanishes. In recent decades, modern results obtained from many physical models indicate that particle dynamics can be influenced by the energy regime under which they are tested. In this context,  the so-called doubly special relativity theories include possible modifications of the energy-momentum relations that, in principle, can be tested in a number of observations including ultra high energy cosmic
rays \cite{rainbow3}. Such modifications of the dispersion relations of special relativity can be extended for the case of curved spaces in the so-called gravity's rainbow \cite{rainbow1}. In this theory, the Planck energy $E_p=\sqrt{\hbar c^5/G }$ is an important parameter that can be used to separate the classical from quantum behavior of physical objects. 

The dynamics of scalar and fermionic particles in curved spaces has been studied in a variety of contexts \cite{santos1,santos2,santos6,bakke1,bakke2,bakke4,bakke5,Barros1,barros3,vol3,santos3,vol6,vol7,vol8,vol9}, where we can highlight the case of the Klein-Gordon (KG) oscillator under the influence of noninertial effects \cite{santos5,santos7}, in which the authors present two different classes of solutions for the KG equation with vector and scalar potentials in the space-time of a cosmic string. Scalar particles can also be studied using the Duffin--Kemmer--Petiau (DKP) formalism in which a specific representation of the DKP equation allows us to access the scalar sector and obtain information about scalar particles as was done in \cite{Castro1,Castro2} where KG oscillator in a space-time with a topological defect is considered. Other studies of scalar particles involve the Coulomb-type potential in a space-time with a screw dislocation \cite{vol10}, dynamics of the particles in G\"{o}del-type metrics \cite{vol11} and the Dirac oscillator under the presence of the Aharonov--Casher effect \cite{vol12}. Topological defects also play an important role in  hyperbolic metamaterials where in certain cases their behavior can be characterized by the disclination \cite{mmaterial1}. Applications of topological defects in condensed matter can be found, among other factors, due to the well-known analogy between disclinations in solids and cosmic strings \cite{defects}. 

In the scenario of gravity's rainbow, the Landau levels \cite{rainbow10} and the 	
Dirac oscillator  \cite{rainbow9} in the presence of a cosmic string have been studied.  In the present paper, we will consider scalar particles in a cosmic string space-time in the context of gravity's rainbow and obtain the solutions of the KG equation, with both minimum (Coulomb-like)  and non-minimum coupling (KG oscillator) in two scenarios of the gravity's rainbow, and analyze their consequences.

The paper is structured as follows: In Section \ref{sec2} we discuss theories of gravity with modified dispersion relations and obtain the cosmic string line element in the framework of gravity's rainbow. The KG equation in a generalized form, the KG oscillator, the solution of radial equation and energy levels in two scenarios of gravity's rainbow are considered in Section \ref{sec3}. Similarly, in Section \ref{sec4} we consider a Coulomb-like potential in the KG equation and find the solution to the differential equation as well as we obtain the energy spectrum associated with this system. Finally, in Section \ref{sec5} we describe our conclusions.

 \section{Cosmic strings in the context of gravity's rainbow }
 \label{sec2}
 Cosmic strings are extended objects which may have been formed in the early universe during a series of symmetry breaking phase transitions, as proposed by Kibble \cite{string15}. In the 1990's it was believed that they have influenced the observed distribution of matter in the universe and in this way, contributed to the high concentration of matter in galaxy superclusters, nowadays, seems to have a similar role in the context of superstring theory. In the context of GR and modified theories of gravity as the gravity's rainbow, cosmic strings have been studied extensively in the last years. The line element of a cosmic string (in units $c=\hbar =G=1$) in the context of the GR is given by 
 \begin{equation}
     ds^{2}=-dt^{2}+dr^{2}+r^{2}\alpha^{2}d\phi^{2}+dz^{2}
     \label{eq1},
 \end{equation}
 where $\mu$ is the linear mass density of the string. The range of the coordinates are written as $r\in (0,\infty)$, $z\in (-\infty,\infty)$, $\phi \in [0,2\pi)$. The parameter $\alpha = 1-4G\mu$ is related to the angular deficit of the space-time, and the flat space-time is obtained when $\alpha=1$. Here we are interested in the case $0<\alpha<1$ in which the space-time around the string can be associated with a relative deflection of matter or light. Solutions of wave equations in the background of the line element (\ref{eq1}) have been studied extensively, and a striking feature associated with such solutions is related to the energy spectrum. For instance, the set of discrete energies obtained from the Dirac equation with vector and scalar potentials of the Coulomb types under the influence of noninertial effects reported in \cite{santos5} has the form
 \begin{equation}
     \epsilon=-\omega |l| + \frac{e\kappa m \eta}{\zeta + e^2 \kappa^2}\pm \sqrt{\zeta \frac{(p_z^2 + m^2)(\zeta + e^2\kappa^2)-m^2 \eta^2}{(\zeta + e^2\kappa^2)^2}},
     \label{eq1b}
 \end{equation}
where $\zeta=\left(N+\frac{1}{2}+\sqrt{\frac{l^2}{\alpha^2} + \eta^2-e^2 \kappa^2}\right)^2$ with $N$, $e\kappa$, $m$, $\omega$, $\eta$, being the radial quantum number, electromagnetic coupling, particle mass, angular velocity of the rotating frame, and scalar potential parameter, respectively. We can see that the energy of the particle is influenced by $\alpha$ so that Eq. (\ref{eq1b}) provides information about the topology of the space-time around the particle.

 On the other hand, it may be interesting to study such topological defects in modified theories of gravity and verify if these theories may reveal new physical characteristics of space-time. In the case of doubly special relativity theories, the usual dispersion relations may be altered due to deformations of the Lorentz symmetry. Initially, we consider a deformation in the general form
 \begin{equation}
E^2f(E/E_P)^2-p^2g(E/E_P)^2=M^2,
\label{eq2}
\end{equation}
 where $p$, $E$, and $E_P$ are the modulus of momentum, energy of the particle and the Planck energy respectively. In the literature, the choice of rainbow functions $f(E/E_P)$ and $g(E/E_P)$ is theoretically and phenomenologically motivated and they must satisfy the equations 
 \begin{equation}
    \lim_{E/E_P\rightarrow 0}f(E/E_P)=\lim_{E/E_P\rightarrow 0}g(E/E_P)=1,
     \label{eq3}
 \end{equation}
 i.e., in the infrared limit the standard energy--momentum dispersion
relation is recovered. In gravity's rainbow the deformation of the dispersion relation can be incorporated by considering the following energy dependent metric \cite{rainbow1}
\begin{equation}
    g(x)=\eta^{ab} e_{a}(x)\otimes e_{b}(x),
    \label{eq3b}
\end{equation}
where the energy dependent frame fields $e_{a}(x)$ are related to the energy independent frame fields, denoted by $\widetilde{e_{a}}$, as follows:
\begin{equation}
    e_{0}(x)=\frac{1}{\Xi (x)} \widetilde{e_{0}}, \quad e_{i}(x)=\frac{1}{\Sigma (x)} \widetilde{e_{i}}.
    \label{eq3c}
\end{equation}
 
 In this paper we are interested in the following particular solution of the Einstein equations in the framework of gravity's rainbow \cite{rainbow12}
 \begin{equation}
     ds^{2}=-\frac{dt^{2}}{f(x)^{2}}+\frac{1}{g(x)^{2}}(dr^{2}+r^{2}\alpha^{2}d\phi^{2}+dz^{2})
     \label{eq4},
 \end{equation}
 where $x=E/E_P$. It is easy to see that in the limiting case where $f(x)=g(x)=1$, Eq. (\ref{eq4}) reduces to the metric of the cosmic string in GR. This metric has elements of the usual cosmic string space-time encoded by $\alpha$, besides the functions $f$ and $g$ that add the contribution of modified gravity.
 
 \section{Klein-Gordon oscillator in the presence of a cosmic string in the context of gravity's rainbow}
\label{sec3}
Firstly, let us write the covariant equation which will describe scalar particles in relativistic quantum mechanics. Namely the covariant KG equation for a scalar particle of mass $M$ in a curved background, which experiences external vectors $A_\mu$ and $X_\mu$ and a scalar potential $V_s$, reads
\begin{equation}
\left[-\frac{1}{\sqrt{-g}}D_\mu^{(+)}g^{\mu\nu}\sqrt{-g}D_\nu^{(-)}+(M+V_s)^2\right]\psi=0,
    \label{eq5}
\end{equation}
where $D^{(\pm)}=\partial_\mu\pm X_{\mu}+iA_\mu$. Notice that we are assuming that the vector $X_\mu$ associated to the KG oscillator is not minimally coupled in contrast to the electromagnetic potential $A_{\mu}$. Besides that, the potentials in Eq. (\ref{eq5}) behave differently under well-known operations as the charge conjugation. For example, $X_\mu$ and $V_s$ do not distinguish particles from antiparticles while $A_{\mu}$ couple to the charge of the particle. The KG equation has been studied in a series of background space-times and in many cases no analytical solution has been found yet. Its solution in the context of quantum field theory is the first step in a canonical quantization scheme. Let us present in what follows an analytical study of Eq. (\ref{eq5})  and describe four exact solutions in the context of a cosmic string in the scenario of gravity's rainbow.

 Now consider the case of a KG oscillator potential. Taking into account that the space-time is generated by a cosmic string in the context of gravity's rainbow, let us write the KG equation for the KG oscillator potential by setting $A_\mu=0$ and $X_\mu=(0,M\Omega r,0,0)$ in Eq. (\ref{eq5}). Here, due to the presence of the rainbow functions in the line element (\ref{eq4}), the wave equation has a dependence of the energy in a more complicated way. Indeed, we observe that the determinant of the metric corresponding to the space-time (\ref{eq4}) is written as $\sqrt{-g}=\frac{\alpha r}{f(x) g(x)^3}$ and the contravariant form of the metric is given by $g^{\mu\nu}=diag(-f(x)^2,g(x)^2,g(x)^2/r^2\alpha^2,g(x)^2)$. By using these results in Eq. (\ref{eq5}), we obtain the following partial diferential  equation
 
 \begin{equation}
\left[f(x)^2\partial_t^2 - \frac{g(x)^2}{r}(\partial_r + M\Omega r)r(\partial_r - M\Omega r) - \frac{g(x)^2}{r^2\alpha^2}\partial_\phi^2 - g(x)^2\partial_z^2 + M^2 \right]\psi = 0.
     \label{eq6}
 \end{equation}
 Note that the Hamiltonian associated to this expression is independent of the coordinates $t,\phi$ and $z$. Thus it is reasonable to make an ansatz where the solutions which depend on $t,\phi$ and $z$ are given by an exponential form written in what follows and the solution of radial coordinate is a function to be determined, i.e, 
 \begin{equation}
 \psi(t,r,z,\phi) = e^{(-iEt+il\phi+ik_z z)}H(r)/\sqrt{r},
     \label{eq7}
 \end{equation}
 with $E$ being the energy of the particle, $l=0,\pm 1,\pm 2,...$ and $k_z$ is the momentum in the z direction. In fact, it is possible to show that the periodic condition in the angular coordinate imply that $l \in \mathbb{Z}$. Substituting Eq. (\ref{eq7}) into Eq. (\ref{eq6}) and then simplifying terms, we get
 
 \begin{equation}
    \left[\frac{d^2}{dr^2} - M^2\Omega^2 r^2 - \frac{\left(l^2/\alpha^2 - 1/4\right)}{r^2} + K^2 \right]H(r)=0,
     \label{eq8}
 \end{equation}
where $K=\sqrt{\left(\frac{f(x)E}{g(x)}\right)^2 -  k_z^2 - \frac{M^2}{g(x)^2} - 2M\Omega}$. This second order radial differential equation  describes the KG oscillator in a space-time of a cosmic string in the context of the gravity's rainbow. To obtain normalizable eigenfunctions of Eq. (\ref{eq8}), we propose the usual transformation in the radial coordinate $\rho = M\Omega r^2$ and an ansatz in the following way:
\begin{equation}
H(r)=r^{\frac{|l|}{\alpha}+1/2}e^{-\frac{M\Omega r^2}{2}}F(r).
    \label{eq9}
\end{equation}
Then, we obtain the radial equation
\begin{equation}
\rho\frac{d^2F(\rho)}{d\rho^2} + \left(\frac{|l|}{\alpha} + 1 - \rho \right)\frac{dF(\rho)}{d\rho} - \left(\frac{|l|}{2\alpha} + \frac{1}{2} - \frac{K^2}{4M\Omega}\right)F(\rho) = 0,
    \label{eq10}
\end{equation}
that is the confluent hypergeometric equation, a second order linear homogeneous differential equation with two independent solutions . The solution of Eq. (\ref{eq10}), regular at $\rho$ = 0, may be obtained considering the confluent hypergeometric function, denoted by $F(\rho)=_{1}F_{1}(a,b;\rho)$
so that 
\begin{align}
    a &= \frac{|l|}{2\alpha} + \frac{1}{2} - \frac{K^2}{4M\Omega}, \label{eq11}\\
    b &=  \frac{|l|}{\alpha} + 1 ,
    \label{eq12}\\
   \rho &= M\Omega r^2.
   \label{eq12b}
\end{align}
 In fact, the most general solution of Eq. (\ref{eq10}) equation is given by the combination 
 \begin{equation}
 F(\rho)=A\:\: _{1}F_{1}(a,b;\rho) + B\rho^{1-b}\:\: _{1}F_{1}(a-b,2-b;\rho),
     \label{eq13}
 \end{equation}
where $A$ and $B$ are arbitrary constants. As the second term of Eq. (\ref{eq13}) is singular at the
origin, i.e., $\rho = 0$ , we impose that $B=0$. An important feature of the confluent hypergeometric function, related to its asymptotic behavior \cite{abramo,santos7}, demands that the parameter $a$ to be a negative integer. This is a well known result that permit us obtain with the energy spectrum of this kind of particle. Now, to study the energy spectrum in detail, we must choose the shape of the rainbow functions. 
 
 \subsection{First Case:}
\begin{figure}[H]
 \centering
\includegraphics[scale=0.5]{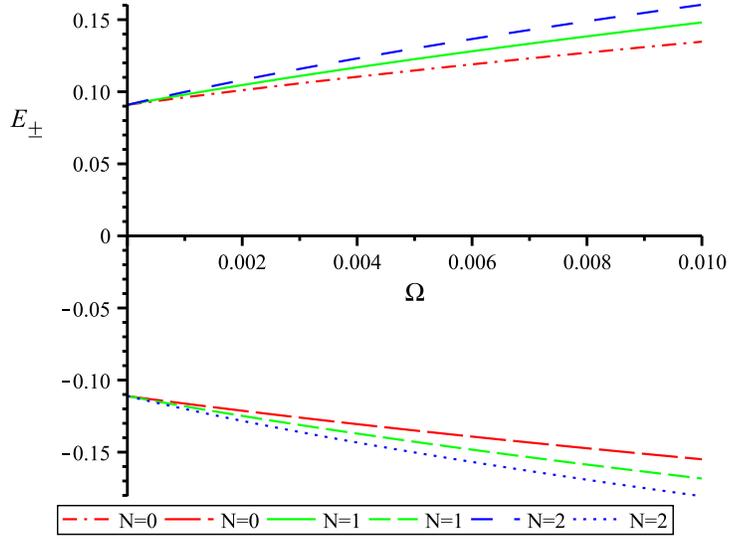}\newline
\caption{Energy as a function of $\Omega$ in which $E_{\pm}$ is plotted for three different values of the quantum number $N$ defined in Eq. (\ref{eq14}) with
parameters $\alpha=0.6$, $M=0.1$, $l=2$
and $k_{z}=0.$}
\label{fig1}
\end{figure}

\begin{figure}[H]
 \centering
\includegraphics[scale=0.5]{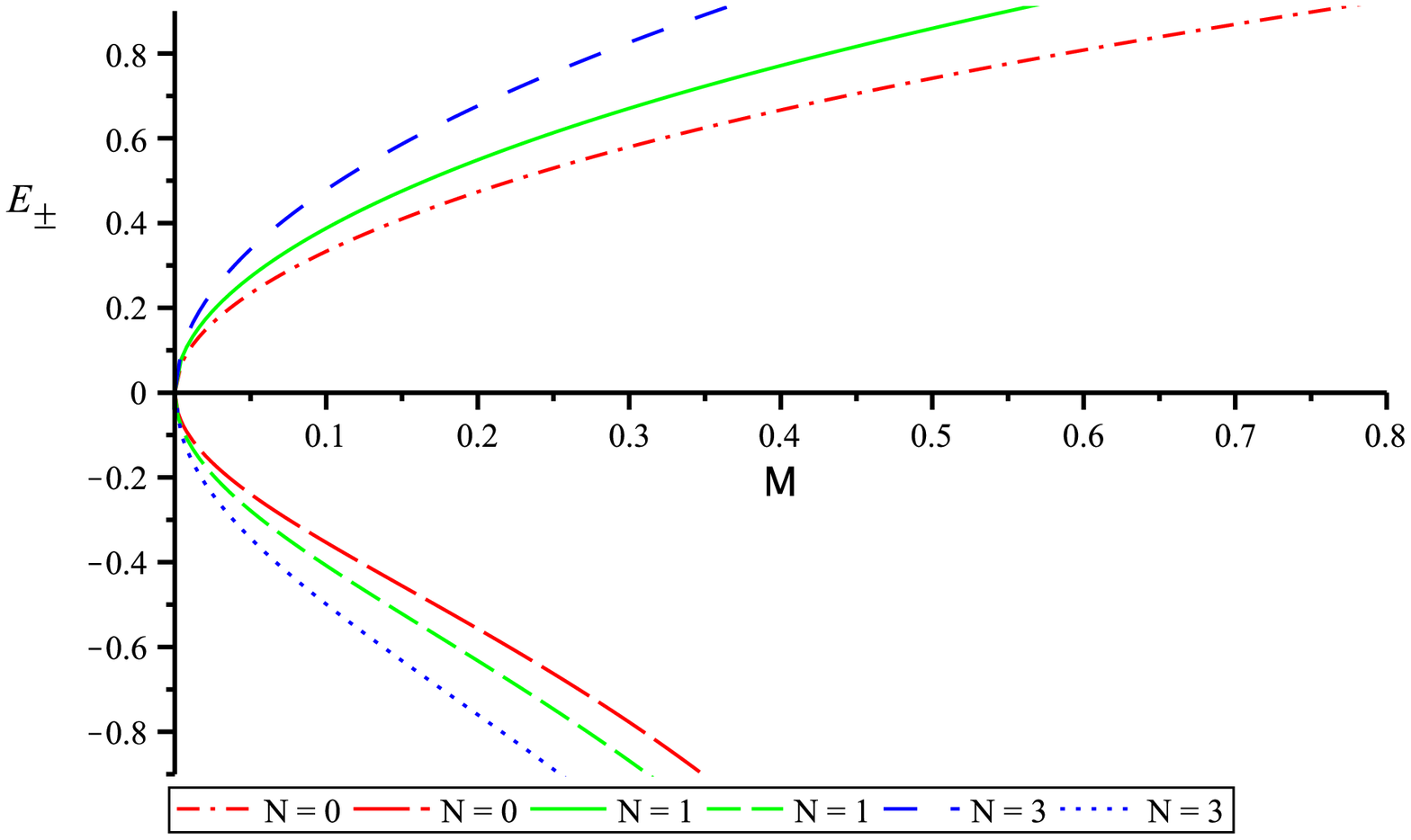}\newline
\caption{Energy as a function of $M$ in which $E_{\pm}$ is plotted for three different values of the quantum number $N$ with
parameters $\alpha=0.6$, $\Omega=0.1$, $l=2$
and $k_{z}=0.$}
\label{fig2}
\end{figure}

\begin{figure}[H]
 \centering
\includegraphics[scale=0.6]{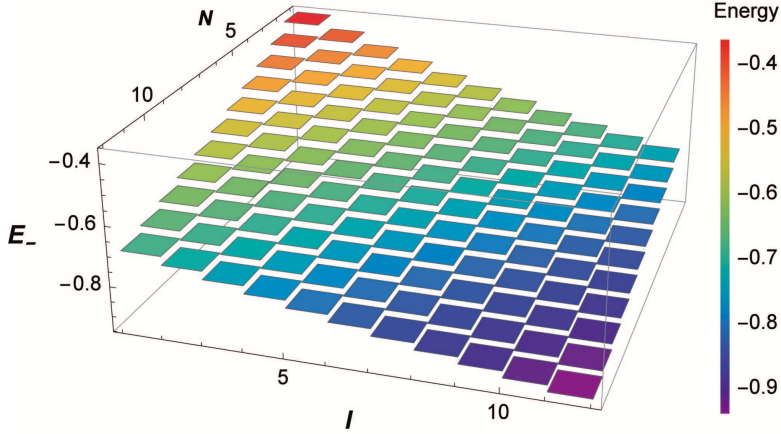}\newline
\caption{Energy as function of $N$ and $l$ in which $E_{-}$ is plotted with
parameters $\alpha=0.6$, $M=0.1$, $\Omega =2$
and $k_{z}=0.$}
\label{fige1}
\end{figure}

As we have discussed before, the parameter $a$ in the hypergeometric function must be a negative integer, namely  $a=-N$ where $N=0,\pm 1,\pm 2,...$ (see  \cite{santos7} for more details). In this way, by using Eq. (\ref{eq11}) and the definition of the parameter $K$, we obtain the quantization condition
\begin{equation}
 \frac{1}{2}\frac{|l|}{\alpha} - \frac{f(x)^2E^2}{4g(x)^2M\Omega} + \frac{k_z^2}{4M\Omega} + \frac{M}{4g(x)^2\Omega} + 1 = -N.
\label{eq14}
\end{equation}
The above equation can be solved exactly depending on the choice of the rainbow functions. Several choices of these functions are proposed in the literature and for the purpose of this paper, we consider initially the following case
 \begin{equation}
 \label{eq16}
f(x)=g(x)=\frac{1}{1-x},
 \end{equation}
where the temporal and the spatial function have the same behavior. This function was studied in \cite{rainbow2}, and it provides a deformation of the dispersion relation where the speed of light is constant. By substituting Eq. (\ref{eq16}) into Eq. (\ref{eq14}) and solving the expression for $E$, we can write the energy spectrum in the form
 \begin{equation}
 E_{\pm} = \frac{\frac{M^2}{E_p} \pm \sqrt{4M\Omega\xi_{Nlk_z} -\frac{4\Omega M^3\xi_{Nlk_z}}{E_p^2} +M^2}}{\frac{M^2}{E_p^2}-1},   
 \label{eq17}
 \end{equation}
where $\xi_{N l k_z}=N+\frac{|l|}{2\alpha}+\frac{k_z^2}{4M\Omega}+1$ with $N=0, 1, 2,...$ and so on. Remarkably, the use of functions (\ref{eq16}) break the energy symmetry of KG oscillator around $E$. As we can see in Eq. (\ref{eq17}), the term $M^2/E_p$ is associated with such symmetry breaking. Figures \ref{fig1} and \ref{fig2} show the energy as a function of frequency and of the mass, respectively. It is possible to see clearly that both graphs are not symmetrical with respect to the horizontal axis. The energy as a function of the variables $N$ and
$l$ are shown in the plot of Fig. \ref{fige1}.

\subsection{Second Case:}
\begin{figure}[H]
\centering
\includegraphics[scale=0.5]{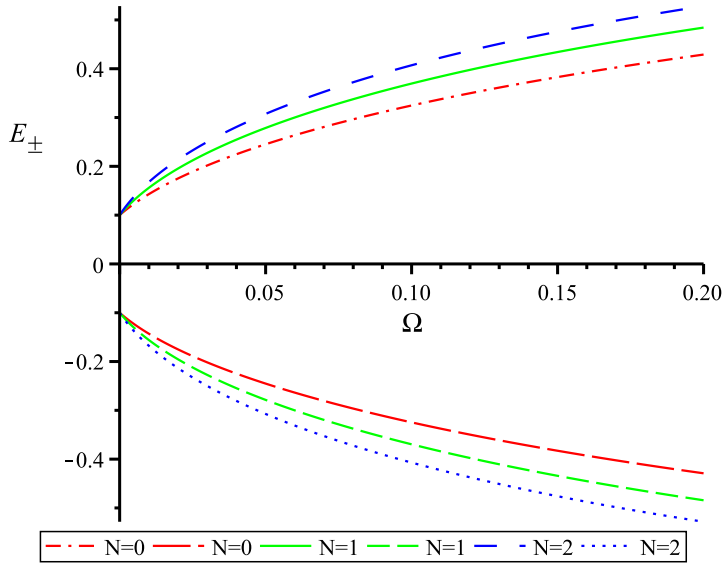}\newline
\caption{Energy as a function of $\Omega$ in which $E_{\pm}$ is plotted for three different values of the quantum number $N$ defined in Eq. (\ref{eq14}) with
parameters $\alpha=0.6$, $M=0.1$, $l=2$
and $k_{z}=0.$}
\label{fig3}
\end{figure}

\begin{figure}[H]
 \centering
\includegraphics[scale=0.5]{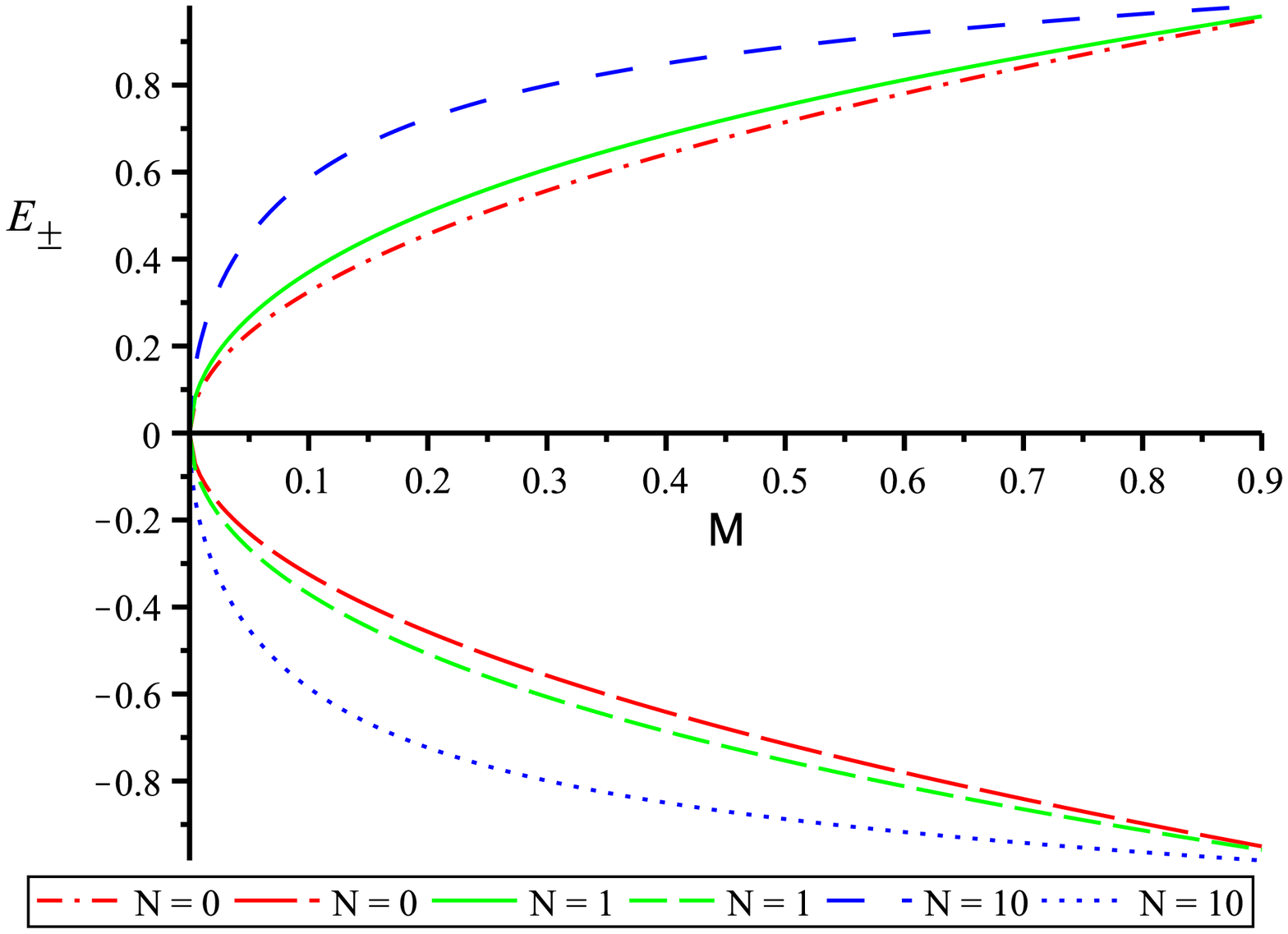}\newline
\caption{Energy as a function of $M$ in which $E_{\pm}$ is plotted for three different values of the quantum number $N$ with
parameters $\alpha=0.6$, $\Omega=0.1$, $l=2$
and $k_{z}=0.$}
\label{fig4}
\end{figure}

\begin{figure}[H]
 \centering
\includegraphics[scale=0.6]{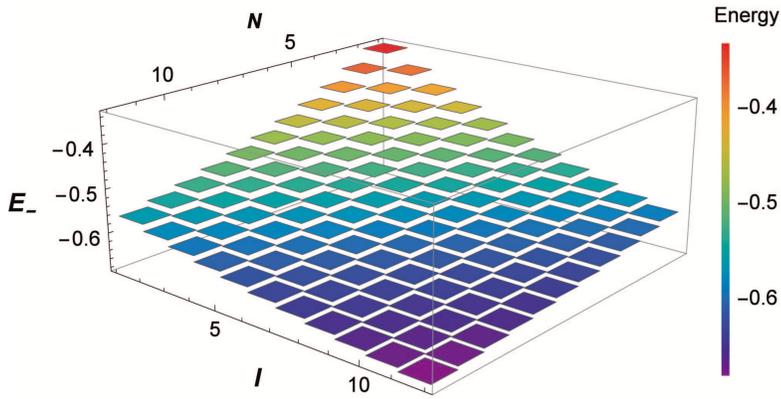}\newline
\caption{Energy as function of $N$ and $l$ in which $E_{-}$ is plotted with
parameters $\alpha=0.6$, $M=0.1$, $\Omega =2$
and $k_{z}=0.$}
\label{fige2}
\end{figure}

We can see that the KG Oscillator in the context of gravity's rainbow has a rich structure that generalizes the results of the usual Oscillator in the GR. Another possibility is to consider a different choice for the rainbow function and compare with the results previously obtained. Here, the second choice is
\begin{equation}
f(x)=1;\:\:\:g(x)=\sqrt{1-x^2}.
\label{eq15}
\end{equation}
This rainbow function prevents the catastrophe of the Hawking radiation when the black hole mass decreases \cite{rainbow27}. By substituting this function into Eq. (\ref{eq14}) and solving the obtained expression for $E$, we can write the energy spectrum for the KG oscillator as
\begin{equation}
\label{eq16b}
E_{\pm}=\pm\sqrt{\frac{4\Omega\xi_{N l k_z} M+M^2}{\frac{4\Omega \xi_{N l k_z} M}{E_p^2} + 1}},    
\end{equation}
where $\xi_{N l k_z}=N+\frac{|l|}{2\alpha}+\frac{k_z^2}{4M\Omega}+1$. This is the energy spectrum associated with scalar particles under the effect of a space-time generated by a cosmic string and taking into account a dispersion relation deformed due to the gravity's rainbow. Eq. (\ref{eq16b}) has new features which emerge from the combination of the topology of the space-time and of the dependence on the Planck energy. First of all, we can analyze the behavior of the energy spectrum as a function of the oscillator frequency. In Figures \ref{fig3} and \ref{fig4}, we can see that, in absolute values, $E$ tends to increase as $\Omega$ increases. This is an expected result since the increased oscillation frequency is associated with the energy supplied to the system. As it may be seen in Fig. \ref{fige2}, the spectrum can be ploted as a three-dimensional graph to show $E$ as a function of $N$ and $l$. Note in Figures \ref{fig3} and \ref{fig4}  that Eq. (\ref{eq16b}) is symmetric about the point $E=0$ and, moreover, depends on the deficit angular parameter $\alpha$. Another important result is obtained when $E/E_p \longrightarrow 0 $ where the usual KG Oscillator in GR is recovered. 

\section{Coulomb-Like vector potential in the presence of a cosmic string in the context of gravity's rainbow }
\label{sec4}
At this point, we study the influence of the topological features of the cosmic string space-time in gravity's rainbow on scalar particles which experiences a Coulomb-like potential. This potential, in the usual space-time of a cosmic string in GR, allows the formation of bound states. Here we want to find such states and show how they are related to the choices of rainbow functions. By considering the potential $A_\mu=(-a/r,0,0,0)$, the determinant  $\sqrt{-g}=\frac{\alpha r}{f(x)g(x)^3}$ and the contravariant form of the metric is given by
\begin{equation}
\label{eq18}
g^{\mu\nu}=diag(-f(x)^2,g(x)^2,g(x)^2/r^2\alpha^2,g(x)^2),    
\end{equation}
and then Eq. (\ref{eq5}) becomes,
\begin{equation}
\label{eq19}
 \left[f(x)^2(\partial_t + iA_0)(\partial_t + iA_0) - \frac{g(x)^2}{r}\partial_{r}r\partial_r  - \frac{g(x)^2}{\alpha^2 r^2}\partial_{\phi}^2 - g(x)^2\partial^2_z + M^2\right]\psi = 0.
\end{equation}
Similarly to the case of the previous section, the Hamiltonian associated with this equation does not depend explicitly on the coordinates $t,\phi$ and $z$, and thus it is possible to propose a solution in the form
\begin{equation}
\psi(t,r,z,\phi)=e^{(-iEt+il\phi+ik_z z)}G(r).
    \label{eq19b}
\end{equation}
Substituting Eq. (\ref{eq19b}) into Eq. (\ref{eq19}), and simplifying terms, we get the following equation for the radial function $G(r)$
\begin{equation}
\left[\label{eq20}
\frac{d^2}{dr^2} + \frac{1}{r}\frac{d}{dr} - \frac{D^2}{r^2} - \frac{2Z}{r} - C^2 \right]G(r) = 0,
\end{equation}
where 
\begin{equation}
\label{eq21}
D^2 = \frac{l^2}{\alpha^2} - \frac{a^2f(x)^2}{g(x)^2},\:\:\:\:Z = -\frac{aE f(x)^2}{g(x)^2},\:\:\:\:C^2=\frac{M^2}{g(x)^2} + k_z^2 - \frac{f(x)^2 E^2}{g(x)^2}.
\end{equation}
Bound states can be obtained for $Z<0$ which implies
that $E.a>0$. We can see from Eq. (\ref{eq21}) that the effect of gravity's rainbow is codified in the constant parameters $D,Z$ and $C$, which unlike the usual KG oscillator in GR, depends on the rainbow functions $f(x)$ and $g(x)$ as expected. Due to the fact that the parameter $x$ in the rainbow functions depends on the energy of the particle, in order to simplify the computation of the energy levels, it is necessary to impose some restrictions on the class of rainbow functions. 

Now, considering a transformation of the radial coordinate $\rho = 2Cr$, the normalizable solution may be written as
\begin{equation}
\label{eq22}
 G(\rho) = \rho^D e^{-\frac{\rho}{2}}H(\rho),
\end{equation}
and then we obtain the following differential equation for $H(\rho)$ 
\begin{equation}
\rho\frac{d^2 H}{d\rho^2} + (2D+1-\rho)\frac{dH}{d\rho}+ \left(-D-\frac{Z}{C}-\frac{1}{2}\right)H = 0,
\label{eq23}
\end{equation}
which is a confluent hipergeometric function whose solution is 
\begin{align}
        H(\rho) &= _{1}F_{1}(A,B;\rho),  \label{eq24} 
\end{align}
with A and B given by
\begin{align}
     A &= D+\frac{Z}{C} + \frac{1}{2}, \label{eq25}\\
        B &= 2D + 1
        \label{eq26}.
\end{align}

The physical parameters of the confluent hypergeometric function (\ref{eq23}) do not appear to be a trivial combination of constants and should be carefully analyzed. To get an accurate picture of the radial equation, we must choose appropriately the rainbow functions, as was done in previous sections. For this purpose we will choose the same rainbow functions used previously in order to systematize the analysis of results.

\subsection{First Case:}
As it has been done before, we are going to test different choices for the functions that deform the space-time dispersion relation, that is, the rainbow functions. As we will see, in the case of the Coulomb type potential, the choice of $f(x)$ and $g(x)$ with identical form simplifies the analysis of the energy spectrum. Indeed, in the case where $f(x)$ differs from $g(x)$, the quantization condition gives a result that cannot be analytically solved. We can start by analyzing the problem for the potential which depends on the radial coordinate with the following rainbow functions
\begin{equation}
\label{eq27}
f(x)=g(x)=\frac{1}{1-x}.
\end{equation}
In this case, the condition $Z<0$ implies $a>0$ and $E<0$ or  $a<0$ and $E>0$. Additionally, from condition $C^2>0$, we obtain that bound states are expected for $\varepsilon_-<E<\varepsilon_+$, where 

\begin{equation}
    \label{eeqq1}
    \varepsilon_\pm = \frac{-M^2}{E_p\left(1-\frac{M^2}{E_p^2}\right)} \pm \frac{\sqrt{M^2+k_z^2 \left(1-\frac{M^2}{E_p^2}\right)}}{\left(1-\frac{M^2}{E_p^2}\right)}
\end{equation}

\begin{figure}[H]
 \centering
\includegraphics[scale=0.5]{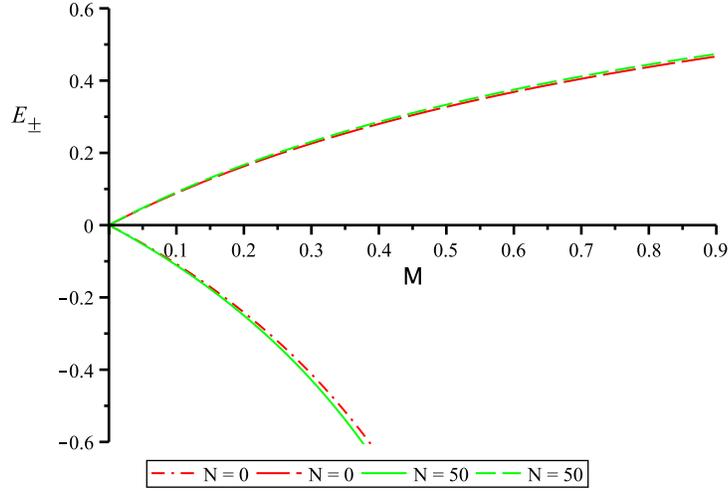}\newline
\caption{Energy as a function of $M$ in which $E_{\pm}$ is plotted for two different values of the quantum number $N$ with
parameters $\alpha=0.6$, $a=\pm 0.5$, $l=2$
and $k_{z}=0.$}
\label{fig5}
\end{figure}

\begin{figure}[H]
 \centering
\includegraphics[scale=1]{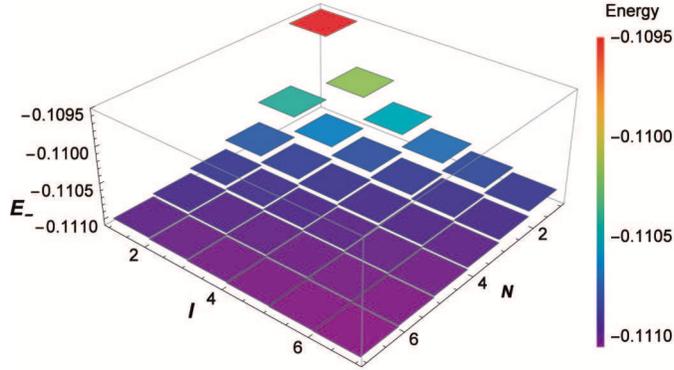}\newline
\caption{Energy as function of $N$ and $l$ in which $E_{-}$ is plotted with
parameters $\alpha=0.6$, $M=0.1$, $a =0.5$
and $k_{z}=0.$}
\label{fige3}
\end{figure}

Therefore, we can conclude that bound-states solutions have a finite set of discrete energy in the range $\varepsilon_-<E<0$ for $a>0$ (repulsive interaction) or $0<E<\varepsilon_+$ for $a<0$ (attractive interaction).  We can see that the rainbow function (\ref{eq27}) has shifted the rest energy of the particles, and the bound states are found in a range of energy that differs from the usual case in GR. The the condition $A=-N$, previously discussed, can be used to derive the following equation
\begin{equation}
\label{eq28}
D + \frac{Z}{C}+\frac{1}{2} = -N.
\end{equation}
This equation can be solved for $E$ and thus providing the energy of the particle. To this end, we use Eq. (\ref{eq27}) and the definitions of $C$, $D$ and $Z$. The result can be written as
\begin{equation}
\label{eq29}
E_{\pm} = \frac{\frac{-M^2\xi}{E_p}}{a^2+\xi - \frac{M^2\xi}{E_p^2}} \pm \frac{\sqrt{\xi}\sqrt{(k_z^2+M^2)(\xi + a^2) - \frac{ k_z^2 M^2\xi}{E_p^2}}}{a^2+\xi - \frac{M^2\xi}{E_p^2}},
\end{equation}
where $\xi=\left(N + 1/2 + \sqrt{\frac{l^2}{\alpha^2} - a^2}\right)^2$. Again, we see an expression that is much more complex than the usual results in GR. The first term in the right side of Eq. (\ref{eq29}) breaks the symmetry around the point $E=0$, i.e., the energies of antiparticles is greater than the one corresponding to particles, in absolute values. In Fig. \ref{fig5}, we observe that the plot is not symmetrical around the horizontal axis, as expected. On the other hand, the spectrum can be plotted as a three-dimensional graph to show $E$ as a function of $N$ and $l$, as Fig. \ref{fige3} illustrates.

\subsection{Second Case:}
The rainbow function in the second scenario corresponds to the following choice
\begin{equation}
f(x)=1;\:\:\:g(x)=\sqrt{1-x^2}.
\label{eq30}
\end{equation}
In this case the condition $Z<0$ again implies $a>0$ and $E<0$ or $a<0$ and $E>0$. From condition $C^2>0$ we obtain that bound states are expected for $\varepsilon_-<E<\varepsilon_+$, where 

\begin{equation}
    \label{extra1}
    \varepsilon_\pm = \pm \frac{E_p}{E_p^2+k_z^2}\sqrt{(E_p^2+k_z^2)(k_z^2+M^2)}.
\end{equation}

Therefore, we can conclude that bound states solutions have a finite set of discrete energies in the range  $\varepsilon_-<E<0$ for $a>0$ (repulsive interaction) or $0<E<\varepsilon_+$ for $a<0$ (attractive interaction). The functions $ f (x) $ and $ g (x) $ have different forms. As a consequence we have no canceling terms as in the previous problem and thus the spectrum cannot be solved in a simple way. Anyway, we will do a numerical analysis of the energy spectrum associated with this problem. By substituting Eq. (\ref{eq30}) into expression (\ref{eq28}), we find that the energy spectrum obeys the relation
\begin{equation}
\label{eq31}
\sqrt{\frac{l^2}{\alpha^2} - \frac{a^2}{1 - \frac{E^2}{E_p^2}}} - \frac{aE}{(1 - \frac{E^2}{E_p^2})\sqrt{k_z^2 + \frac{M^2 - E^2}{1 - \frac{E^2}{E_p^2}}}} + \frac{1}{2} = -N.
\end{equation}
The first numerical values of $N$ and their respective energies are listed in Table 1. We can see that the energy is symmetrical for the rainbow functions (\ref{eq30}). This behavior is similar to the second scenario studied in the case of the KG oscillator. It is important to point out that Eq. (\ref{eq31}) can be solved with a first order expansion in $E/Ep$, and the results obtained  by this approach are similar to the numerical values, shown in the Table 1.
\begin{table}[H]
\label{tab1}
\caption{The first values of $N$ \ that satisfy the
quantization condition for $a=0.5$ (Negative energies) and for $a=-0.5$ (Positive energies).} 
\begin{ruledtabular}
\begin{tabular}{cccc}
$l$ & $\alpha$ & $N$  & $E_{\pm}$   \tabularnewline
\hline
1 & 0.9 &  0  & $\pm$ 0.09476826903    \tabularnewline
1 & 0.9 &  1  & $\pm$ 0.09802605150    \tabularnewline
1 & 0.9 &  2  & $\pm$ 0.09898000068    \tabularnewline
1 & 0.9 &  3  & $\pm$ 0.09937987293    \tabularnewline
1 & 0.9 &  4  & $\pm$ 0.09958387914    \tabularnewline
2 & 0.9 &  0  & $\pm$ 0.09826843410    \tabularnewline
2 & 0.9 &  1  & $\pm$ 0.09907305723    \tabularnewline
2 & 0.9 &  2  & $\pm$ 0.09942479351    \tabularnewline
2 & 0.9 &  3  & $\pm$ 0.09960885437    \tabularnewline
2 & 0.9 &  4  & $\pm$ 0.09971696060    \tabularnewline
3 & 0.6 &  0  & $\pm$ 0.09958141247    \tabularnewline
3 & 0.6 &  1  & $\pm$ 0.09970018406    \tabularnewline
3 & 0.6 &  2  & $\pm$ 0.09977478347    \tabularnewline
3 & 0.6 &  3  & $\pm$ 0.09982467801    \tabularnewline
3 & 0.6 &  4  & $\pm$ 0.09985964820    \tabularnewline
\end{tabular}
\end{ruledtabular}
\end{table}

\section{Conclusions}
\label{sec5}
We have obtained the solutions of the KG equation and studied the energy spectrum for the KG oscillator and a vector potential of the Coulomb type in the context of gravity's rainbow. We have considered two set of rainbow functions where significant features of the space-time may be observed. In the first scenario, we have chosen the functions defined in Eq. (\ref{eq16}) and (\ref{eq15}) which are well studied in the literature.
 In the case of KG oscillator, we have obtained a radial differential equation that may be solved analytically using hypergeometric functions. The energy spectrum associated to rainbow functions (\ref{eq16}) and (\ref{eq15}) is obtained and the results show that the use of function (\ref{eq15}) break the energy symmetry of KG oscillator around $E$. 
 
 In the case of Coulomb-like potential, the differential equations obtained seem to be more complicated than the ones that appeal in the study of the of KG oscillator. Even so, in the first scenario studied we have achieved a radial equation that can be solved using the hypergeometric function, as in the KG oscillator. In both scenarios we have considered bound states in which the condition $Ea>0$ is satisfied. In the second case, we numerically solve the energy spectrum and the result obtained can be seen in the Table 1. The energy values increase along with the quantum numbers and it is also possible to see that the particle and anti-particle spectrum is symmetrical around $E$, as expected for this rainbow function. 
 
 The symmetry breaking of the energy levels of particle and anti-particle pointed out in this work is interesting and can also be observed in the situation where the particle is free. In the case of rainbow functions given by Eq. (\ref{eq15}), we can observe this behavior both for free states and for bound states. As a future perspective, we plan to use others rainbow functions and a class of different types of potentials.  Finally, we have observed that the usual results in GR are recovered at the low energy limit ($E_p \rightarrow \infty$) of the test particle as should be expected.

\section{Acknowledgments}
 C.E. Mota has a scholarship paid by Capes (Brazil) and L.C.N. Santos would like to thank Conselho Nacional de Desenvolvimento Científico e Tecnológico (CNPq) for partial financial 
support through the research Project No. 155361/2018-0.
This work was supported in part by means of funds provided by CNPq, Brazil, Grant No. 307932/2017-6 (PQ) and No. 422755/2018-4 (UNIVERSAL), S\~{a}o Paulo Research Foundation (FAPESP), Grant No. 2018/20577-4 e FAPEMA, Brazil, Grant No. UNIVERSAL-01220/18.

\bibliographystyle{aipnum4-1}
\bibliography{referencias_unificadas}

\begin{thebibliography}{39}%
\makeatletter
\providecommand \@ifxundefined [1]{%
 \@ifx{#1\undefined}
}%
\providecommand \@ifnum [1]{%
 \ifnum #1\expandafter \@firstoftwo
 \else \expandafter \@secondoftwo
 \fi
}%
\providecommand \@ifx [1]{%
 \ifx #1\expandafter \@firstoftwo
 \else \expandafter \@secondoftwo
 \fi
}%
\providecommand \natexlab [1]{#1}%
\providecommand \enquote  [1]{``#1''}%
\providecommand \bibnamefont  [1]{#1}%
\providecommand \bibfnamefont [1]{#1}%
\providecommand \citenamefont [1]{#1}%
\providecommand \href@noop [0]{\@secondoftwo}%
\providecommand \href [0]{\begingroup \@sanitize@url \@href}%
\providecommand \@href[1]{\@@startlink{#1}\@@href}%
\providecommand \@@href[1]{\endgroup#1\@@endlink}%
\providecommand \@sanitize@url [0]{\catcode `\\12\catcode `\$12\catcode
  `\&12\catcode `\#12\catcode `\^12\catcode `\_12\catcode `\%12\relax}%
\providecommand \@@startlink[1]{}%
\providecommand \@@endlink[0]{}%
\providecommand \url  [0]{\begingroup\@sanitize@url \@url }%
\providecommand \@url [1]{\endgroup\@href {#1}{\urlprefix }}%
\providecommand \urlprefix  [0]{URL }%
\providecommand \Eprint [0]{\href }%
\providecommand \doibase [0]{http://dx.doi.org/}%
\providecommand \selectlanguage [0]{\@gobble}%
\providecommand \bibinfo  [0]{\@secondoftwo}%
\providecommand \bibfield  [0]{\@secondoftwo}%
\providecommand \translation [1]{[#1]}%
\providecommand \BibitemOpen [0]{}%
\providecommand \bibitemStop [0]{}%
\providecommand \bibitemNoStop [0]{.\EOS\space}%
\providecommand \EOS [0]{\spacefactor3000\relax}%
\providecommand \BibitemShut  [1]{\csname bibitem#1\endcsname}%
\let\auto@bib@innerbib\@empty
\bibitem [{\citenamefont {Abbott}\ \emph {et~al.}(2016)\citenamefont {Abbott},
  \citenamefont {Abbott}, \citenamefont {Abbott}, \citenamefont {Zhang},
  \citenamefont {Zhao}, \citenamefont {Zhou}, \citenamefont {Zhou},
  \citenamefont {Zhu}, \citenamefont {Zucker}, \citenamefont {Zuraw},\ and\
  \citenamefont {Zweizig}}]{waves1}%
  \BibitemOpen
  \bibfield  {author} {\bibinfo {author} {\bibfnamefont {B.~P.}\ \bibnamefont
  {Abbott}}, \bibinfo {author} {\bibfnamefont {R.}~\bibnamefont {Abbott}},
  \bibinfo {author} {\bibfnamefont {M.}~\bibnamefont {Abbott}}, \bibinfo
  {author} {\bibfnamefont {Y.}~\bibnamefont {Zhang}}, \bibinfo {author}
  {\bibfnamefont {C.}~\bibnamefont {Zhao}}, \bibinfo {author} {\bibfnamefont
  {M.}~\bibnamefont {Zhou}}, \bibinfo {author} {\bibfnamefont {Z.}~\bibnamefont
  {Zhou}}, \bibinfo {author} {\bibfnamefont {X.~J.}\ \bibnamefont {Zhu}},
  \bibinfo {author} {\bibfnamefont {M.~E.}\ \bibnamefont {Zucker}}, \bibinfo
  {author} {\bibfnamefont {S.~E.}\ \bibnamefont {Zuraw}}, \ and\ \bibinfo
  {author} {\bibfnamefont {J.}~\bibnamefont {Zweizig}} (\bibinfo
  {collaboration} {LIGO Scientific Collaboration and Virgo Collaboration}),\
  }\href {\doibase 10.1103/PhysRevLett.116.061102} {\bibfield  {journal}
  {\bibinfo  {journal} {Phys. Rev. Lett.}\ }\textbf {\bibinfo {volume} {116}},\
  \bibinfo {pages} {061102} (\bibinfo {year} {2016})}\BibitemShut {NoStop}%
\bibitem [{\citenamefont {Akiyama}\ \emph {et~al.}(2019)\citenamefont {Akiyama}
  \emph {et~al.}}]{black1}%
  \BibitemOpen
  \bibfield  {author} {\bibinfo {author} {\bibfnamefont {K.}~\bibnamefont
  {Akiyama}} \emph {et~al.} (\bibinfo {collaboration} {Event Horizon
  Telescope}),\ }\href {\doibase 10.3847/2041-8213/ab0ec7} {\bibfield
  {journal} {\bibinfo  {journal} {Astrophys. J.}\ }\textbf {\bibinfo {volume}
  {875}},\ \bibinfo {pages} {L1} (\bibinfo {year} {2019})}\BibitemShut
  {NoStop}%
\bibitem [{\citenamefont {Sotiriou}\ and\ \citenamefont
  {Faraoni}(2010)}]{fder1}%
  \BibitemOpen
  \bibfield  {author} {\bibinfo {author} {\bibfnamefont {T.~P.}\ \bibnamefont
  {Sotiriou}}\ and\ \bibinfo {author} {\bibfnamefont {V.}~\bibnamefont
  {Faraoni}},\ }\href {\doibase 10.1103/RevModPhys.82.451} {\bibfield
  {journal} {\bibinfo  {journal} {Rev. Mod. Phys.}\ }\textbf {\bibinfo {volume}
  {82}},\ \bibinfo {pages} {451} (\bibinfo {year} {2010})},\ \Eprint
  {http://arxiv.org/abs/0805.1726} {arXiv:0805.1726 [gr-qc]} \BibitemShut
  {NoStop}%
\bibitem [{\citenamefont {Mota}\ \emph {et~al.}(2019)\citenamefont {Mota},
  \citenamefont {Santos}, \citenamefont {Grams}, \citenamefont {da~Silva},\
  and\ \citenamefont {Menezes}}]{santos8}%
  \BibitemOpen
  \bibfield  {author} {\bibinfo {author} {\bibfnamefont {C.~E.}\ \bibnamefont
  {Mota}}, \bibinfo {author} {\bibfnamefont {L.~C.~N.}\ \bibnamefont {Santos}},
  \bibinfo {author} {\bibfnamefont {G.}~\bibnamefont {Grams}}, \bibinfo
  {author} {\bibfnamefont {F.~M.}\ \bibnamefont {da~Silva}}, \ and\ \bibinfo
  {author} {\bibfnamefont {D.~P.}\ \bibnamefont {Menezes}},\ }\href {\doibase
  10.1103/PhysRevD.100.024043} {\bibfield  {journal} {\bibinfo  {journal}
  {Phys. Rev. D}\ }\textbf {\bibinfo {volume} {100}},\ \bibinfo {pages}
  {024043} (\bibinfo {year} {2019})}\BibitemShut {NoStop}%
\bibitem [{\citenamefont {Vilenkin}(1981{\natexlab{a}})}]{string2}%
  \BibitemOpen
  \bibfield  {author} {\bibinfo {author} {\bibfnamefont {A.}~\bibnamefont
  {Vilenkin}},\ }\href {\doibase 10.1103/PhysRevD.24.2082} {\bibfield
  {journal} {\bibinfo  {journal} {Phys. Rev. D}\ }\textbf {\bibinfo {volume}
  {24}},\ \bibinfo {pages} {2082} (\bibinfo {year}
  {1981}{\natexlab{a}})}\BibitemShut {NoStop}%
\bibitem [{\citenamefont {Vilenkin}(1981{\natexlab{b}})}]{string3}%
  \BibitemOpen
  \bibfield  {author} {\bibinfo {author} {\bibfnamefont {A.}~\bibnamefont
  {Vilenkin}},\ }\href {\doibase 10.1103/PhysRevD.23.852} {\bibfield  {journal}
  {\bibinfo  {journal} {Phys. Rev. D}\ }\textbf {\bibinfo {volume} {23}},\
  \bibinfo {pages} {852} (\bibinfo {year} {1981}{\natexlab{b}})}\BibitemShut
  {NoStop}%
\bibitem [{\citenamefont {Magueijo}\ and\ \citenamefont
  {Smolin}(2003)}]{rainbow3}%
  \BibitemOpen
  \bibfield  {author} {\bibinfo {author} {\bibfnamefont {J.}~\bibnamefont
  {Magueijo}}\ and\ \bibinfo {author} {\bibfnamefont {L.}~\bibnamefont
  {Smolin}},\ }\href@noop {} {\bibfield  {journal} {\bibinfo  {journal} {Phys.
  Rev. D}\ }\textbf {\bibinfo {volume} {67}},\ \bibinfo {pages} {044017}
  (\bibinfo {year} {2003})}\BibitemShut {NoStop}%
\bibitem [{\citenamefont {Magueijo}\ and\ \citenamefont
  {Smolin}(2004)}]{rainbow1}%
  \BibitemOpen
  \bibfield  {author} {\bibinfo {author} {\bibfnamefont {J.}~\bibnamefont
  {Magueijo}}\ and\ \bibinfo {author} {\bibfnamefont {L.}~\bibnamefont
  {Smolin}},\ }\href@noop {} {\bibfield  {journal} {\bibinfo  {journal} {Class.
  Quant. Grav.}\ }\textbf {\bibinfo {volume} {21}},\ \bibinfo {pages} {1725}
  (\bibinfo {year} {2004})}\BibitemShut {NoStop}%
\bibitem [{\citenamefont {Santos}\ and\ \citenamefont
  {Barros~Jr.}(2016)}]{santos1}%
  \BibitemOpen
  \bibfield  {author} {\bibinfo {author} {\bibfnamefont {L.~C.~N.}\
  \bibnamefont {Santos}}\ and\ \bibinfo {author} {\bibfnamefont {C.~C.}\
  \bibnamefont {Barros~Jr.}},\ }\href@noop {} {\bibfield  {journal} {\bibinfo
  {journal} {Eur. Phys. J. C}\ }\textbf {\bibinfo {volume} {76}},\ \bibinfo
  {pages} {560} (\bibinfo {year} {2016})}\BibitemShut {NoStop}%
\bibitem [{\citenamefont {Santos}\ and\ \citenamefont
  {Barros~Jr.}(2017)}]{santos2}%
  \BibitemOpen
  \bibfield  {author} {\bibinfo {author} {\bibfnamefont {L.~C.~N.}\
  \bibnamefont {Santos}}\ and\ \bibinfo {author} {\bibfnamefont {C.~C.}\
  \bibnamefont {Barros~Jr.}},\ }\href {\doibase 10.1140/epjc/s10052-017-4732-x}
  {\bibfield  {journal} {\bibinfo  {journal} {Eur. Phys. J. C}\ }\textbf
  {\bibinfo {volume} {77}},\ \bibinfo {pages} {186} (\bibinfo {year}
  {2017})}\BibitemShut {NoStop}%
\bibitem [{\citenamefont {Santos}\ and\ \citenamefont
  {Barros~Jr.}(2018{\natexlab{a}})}]{santos6}%
  \BibitemOpen
  \bibfield  {author} {\bibinfo {author} {\bibfnamefont {L.~C.~N.}\
  \bibnamefont {Santos}}\ and\ \bibinfo {author} {\bibfnamefont {C.~C.}\
  \bibnamefont {Barros~Jr.}},\ }\href@noop {} {\bibfield  {journal} {\bibinfo
  {journal} {Int. J. Mod. Phys. A}\ }\textbf {\bibinfo {volume} {33}},\
  \bibinfo {pages} {1850122} (\bibinfo {year}
  {2018}{\natexlab{a}})}\BibitemShut {NoStop}%
\bibitem [{\citenamefont {Bakke}\ and\ \citenamefont {Furtado}(2013)}]{bakke1}%
  \BibitemOpen
  \bibfield  {author} {\bibinfo {author} {\bibfnamefont {K.}~\bibnamefont
  {Bakke}}\ and\ \bibinfo {author} {\bibfnamefont {C.}~\bibnamefont
  {Furtado}},\ }\href {\doibase http://dx.doi.org/10.1016/j.aop.2013.06.007}
  {\bibfield  {journal} {\bibinfo  {journal} {Ann. Phys.}\ }\textbf {\bibinfo
  {volume} {336}},\ \bibinfo {pages} {489 } (\bibinfo {year}
  {2013})}\BibitemShut {NoStop}%
\bibitem [{\citenamefont {Bakke}(2010)}]{bakke2}%
  \BibitemOpen
  \bibfield  {author} {\bibinfo {author} {\bibfnamefont {K.}~\bibnamefont
  {Bakke}},\ }\href {\doibase http://dx.doi.org/10.1016/j.physleta.2010.09.046}
  {\bibfield  {journal} {\bibinfo  {journal} {Phys. Lett. A}\ }\textbf
  {\bibinfo {volume} {374}},\ \bibinfo {pages} {4642 } (\bibinfo {year}
  {2010})}\BibitemShut {NoStop}%
\bibitem [{\citenamefont {Bakke}\ and\ \citenamefont {Furtado}(2009)}]{bakke4}%
  \BibitemOpen
  \bibfield  {author} {\bibinfo {author} {\bibfnamefont {K.}~\bibnamefont
  {Bakke}}\ and\ \bibinfo {author} {\bibfnamefont {C.}~\bibnamefont
  {Furtado}},\ }\href {\doibase 10.1103/PhysRevD.80.024033} {\bibfield
  {journal} {\bibinfo  {journal} {Phys. Rev. D}\ }\textbf {\bibinfo {volume}
  {80}},\ \bibinfo {pages} {024033} (\bibinfo {year} {2009})}\BibitemShut
  {NoStop}%
\bibitem [{\citenamefont {Bakke}\ and\ \citenamefont {Furtado}(2010)}]{bakke5}%
  \BibitemOpen
  \bibfield  {author} {\bibinfo {author} {\bibfnamefont {K.}~\bibnamefont
  {Bakke}}\ and\ \bibinfo {author} {\bibfnamefont {C.}~\bibnamefont
  {Furtado}},\ }\href {\doibase 10.1103/PhysRevD.82.084025} {\bibfield
  {journal} {\bibinfo  {journal} {Phys. Rev. D}\ }\textbf {\bibinfo {volume}
  {82}},\ \bibinfo {pages} {084025} (\bibinfo {year} {2010})}\BibitemShut
  {NoStop}%
\bibitem [{\citenamefont {Barros~Jr.}(2005)}]{Barros1}%
  \BibitemOpen
  \bibfield  {author} {\bibinfo {author} {\bibfnamefont {C.~C.}\ \bibnamefont
  {Barros~Jr.}},\ }\href@noop {} {\bibfield  {journal} {\bibinfo  {journal}
  {Eur. Phys. J. C}\ }\textbf {\bibinfo {volume} {42}},\ \bibinfo {pages} {119}
  (\bibinfo {year} {2005})}\BibitemShut {NoStop}%
\bibitem [{\citenamefont {Barros~Jr.}(2006)}]{barros3}%
  \BibitemOpen
  \bibfield  {author} {\bibinfo {author} {\bibfnamefont {C.~C.}\ \bibnamefont
  {Barros~Jr.}},\ }\href@noop {} {\bibfield  {journal} {\bibinfo  {journal}
  {Eur. Phys. J. C}\ }\textbf {\bibinfo {volume} {45}},\ \bibinfo {pages} {421}
  (\bibinfo {year} {2006})}\BibitemShut {NoStop}%
\bibitem [{\citenamefont {Hassanabadi}\ \emph {et~al.}(2019)\citenamefont
  {Hassanabadi}, \citenamefont {Hosseini}, \citenamefont {Zare},\ and\
  \citenamefont {Hosseinpour}}]{vol3}%
  \BibitemOpen
  \bibfield  {author} {\bibinfo {author} {\bibfnamefont {H.}~\bibnamefont
  {Hassanabadi}}, \bibinfo {author} {\bibfnamefont {M.}~\bibnamefont
  {Hosseini}}, \bibinfo {author} {\bibfnamefont {S.}~\bibnamefont {Zare}}, \
  and\ \bibinfo {author} {\bibfnamefont {M.}~\bibnamefont {Hosseinpour}},\
  }\href@noop {} {\bibfield  {journal} {\bibinfo  {journal} {Few-Body Syst.}\
  }\textbf {\bibinfo {volume} {60}},\ \bibinfo {pages} {12} (\bibinfo {year}
  {2019})}\BibitemShut {NoStop}%
\bibitem [{\citenamefont {Santos}\ and\ \citenamefont
  {Barros~Jr.}(2019)}]{santos3}%
  \BibitemOpen
  \bibfield  {author} {\bibinfo {author} {\bibfnamefont {L.~C.~N.}\
  \bibnamefont {Santos}}\ and\ \bibinfo {author} {\bibfnamefont {C.~C.}\
  \bibnamefont {Barros~Jr.}},\ }\href {\doibase 10.1142/S0219887819501408}
  {\bibfield  {journal} {\bibinfo  {journal} {International Journal of
  Geometric Methods in Modern Physics}\ }\textbf {\bibinfo {volume} {16}},\
  \bibinfo {pages} {1950140} (\bibinfo {year} {2019})},\ \Eprint
  {http://arxiv.org/abs/https://doi.org/10.1142/S0219887819501408}
  {https://doi.org/10.1142/S0219887819501408} \BibitemShut {NoStop}%
\bibitem [{\citenamefont {Hosseini}, \citenamefont {Hassanabadi},\ and\
  \citenamefont {Hassanabadi}(2019)}]{vol6}%
  \BibitemOpen
  \bibfield  {author} {\bibinfo {author} {\bibfnamefont {M.}~\bibnamefont
  {Hosseini}}, \bibinfo {author} {\bibfnamefont {H.}~\bibnamefont
  {Hassanabadi}}, \ and\ \bibinfo {author} {\bibfnamefont {S.}~\bibnamefont
  {Hassanabadi}},\ }\href@noop {} {\bibfield  {journal} {\bibinfo  {journal}
  {Pramana}\ }\textbf {\bibinfo {volume} {93}},\ \bibinfo {pages} {16}
  (\bibinfo {year} {2019})}\BibitemShut {NoStop}%
\bibitem [{\citenamefont {Cadoni}\ and\ \citenamefont {Mignemi}(1995)}]{vol7}%
  \BibitemOpen
  \bibfield  {author} {\bibinfo {author} {\bibfnamefont {M.}~\bibnamefont
  {Cadoni}}\ and\ \bibinfo {author} {\bibfnamefont {S.}~\bibnamefont
  {Mignemi}},\ }\href {\doibase 10.1016/0370-2693(95)01018-L} {\bibfield
  {journal} {\bibinfo  {journal} {Phys. Lett. B}\ }\textbf {\bibinfo {volume}
  {358}},\ \bibinfo {pages} {217} (\bibinfo {year} {1995})},\ \Eprint
  {http://arxiv.org/abs/gr-qc/9505032} {arXiv:gr-qc/9505032 [gr-qc]}
  \BibitemShut {NoStop}%
\bibitem [{\citenamefont {Hosseinpour}, \citenamefont {Hassanabadi},\ and\
  \citenamefont {de~Montigny}(2019)}]{vol8}%
  \BibitemOpen
  \bibfield  {author} {\bibinfo {author} {\bibfnamefont {M.}~\bibnamefont
  {Hosseinpour}}, \bibinfo {author} {\bibfnamefont {H.}~\bibnamefont
  {Hassanabadi}}, \ and\ \bibinfo {author} {\bibfnamefont {M.}~\bibnamefont
  {de~Montigny}},\ }\href {\doibase 10.1140/epjc/s10052-019-6830-4} {\bibfield
  {journal} {\bibinfo  {journal} {Eur. Phys. J. C}\ }\textbf {\bibinfo {volume}
  {79}},\ \bibinfo {pages} {311} (\bibinfo {year} {2019})},\ \Eprint
  {http://arxiv.org/abs/1904.05889} {arXiv:1904.05889 [hep-th]} \BibitemShut
  {NoStop}%
\bibitem [{\citenamefont {Bradler}, \citenamefont {Hayden},\ and\ \citenamefont
  {Panangaden}(2012)}]{vol9}%
  \BibitemOpen
  \bibfield  {author} {\bibinfo {author} {\bibfnamefont {K.}~\bibnamefont
  {Bradler}}, \bibinfo {author} {\bibfnamefont {P.}~\bibnamefont {Hayden}}, \
  and\ \bibinfo {author} {\bibfnamefont {P.}~\bibnamefont {Panangaden}},\
  }\href {\doibase 10.1007/s00220-012-1476-1} {\bibfield  {journal} {\bibinfo
  {journal} {Commun. Math. Phys.}\ }\textbf {\bibinfo {volume} {312}},\
  \bibinfo {pages} {361} (\bibinfo {year} {2012})},\ \Eprint
  {http://arxiv.org/abs/1007.0997} {arXiv:1007.0997 [quant-ph]} \BibitemShut
  {NoStop}%
\bibitem [{\citenamefont {Santos}\ and\ \citenamefont
  {Barros~Jr.}(2018{\natexlab{b}})}]{santos5}%
  \BibitemOpen
  \bibfield  {author} {\bibinfo {author} {\bibfnamefont {L.~C.~N.}\
  \bibnamefont {Santos}}\ and\ \bibinfo {author} {\bibfnamefont {C.~C.}\
  \bibnamefont {Barros~Jr.}},\ }\href {\doibase 10.1140/epjc/s10052-017-5476-3}
  {\bibfield  {journal} {\bibinfo  {journal} {Eur. Phys. J. C}\ }\textbf
  {\bibinfo {volume} {78}},\ \bibinfo {pages} {13} (\bibinfo {year}
  {2018}{\natexlab{b}})}\BibitemShut {NoStop}%
\bibitem [{\citenamefont {Santos}, \citenamefont {Mota},\ and\ \citenamefont
  {Barros~Jr.}(2019)}]{santos7}%
  \BibitemOpen
  \bibfield  {author} {\bibinfo {author} {\bibfnamefont {L.~C.~N.}\
  \bibnamefont {Santos}}, \bibinfo {author} {\bibfnamefont {C.~E.}\
  \bibnamefont {Mota}}, \ and\ \bibinfo {author} {\bibfnamefont {C.~C.}\
  \bibnamefont {Barros~Jr.}},\ }\href@noop {} {\bibfield  {journal} {\bibinfo
  {journal} {Adv. High Energy Phys.}\ }\textbf {\bibinfo {volume} {2019}},\
  \bibinfo {pages} {7} (\bibinfo {year} {2019})}\BibitemShut {NoStop}%
\bibitem [{\citenamefont {Castro}(2016)}]{Castro1}%
  \BibitemOpen
  \bibfield  {author} {\bibinfo {author} {\bibfnamefont {L.~B.}\ \bibnamefont
  {Castro}},\ }\href {\doibase 10.1140/epjc/s10052-016-3904-4} {\bibfield
  {journal} {\bibinfo  {journal} {Eur. Phys. J. C}\ }\textbf {\bibinfo {volume}
  {76}},\ \bibinfo {pages} {1} (\bibinfo {year} {2016})}\BibitemShut {NoStop}%
\bibitem [{\citenamefont {Castro}(2015)}]{Castro2}%
  \BibitemOpen
  \bibfield  {author} {\bibinfo {author} {\bibfnamefont {L.~B.}\ \bibnamefont
  {Castro}},\ }\href {\doibase 10.1140/epjc/s10052-015-3507-5} {\bibfield
  {journal} {\bibinfo  {journal} {Eur. Phys. J. C}\ }\textbf {\bibinfo {volume}
  {75}},\ \bibinfo {pages} {1} (\bibinfo {year} {2015})}\BibitemShut {NoStop}%
\bibitem [{\citenamefont {Vit{\'o}ria}\ and\ \citenamefont
  {Bakke}(2018)}]{vol10}%
  \BibitemOpen
  \bibfield  {author} {\bibinfo {author} {\bibfnamefont {R.~L.~L.}\
  \bibnamefont {Vit{\'o}ria}}\ and\ \bibinfo {author} {\bibfnamefont
  {K.}~\bibnamefont {Bakke}},\ }\href {\doibase 10.1140/epjp/i2018-12310-9}
  {\bibfield  {journal} {\bibinfo  {journal} {Eur. Phys. J. Plus}\ }\textbf
  {\bibinfo {volume} {133}},\ \bibinfo {pages} {490} (\bibinfo {year}
  {2018})}\BibitemShut {NoStop}%
\bibitem [{\citenamefont {Ahmed}(2018)}]{vol11}%
  \BibitemOpen
  \bibfield  {author} {\bibinfo {author} {\bibfnamefont {F.}~\bibnamefont
  {Ahmed}},\ }\href {\doibase 10.1140/epjc/s10052-018-6082-8} {\bibfield
  {journal} {\bibinfo  {journal} {Eur. Phys. J. C}\ }\textbf {\bibinfo {volume}
  {78}},\ \bibinfo {pages} {598} (\bibinfo {year} {2018})}\BibitemShut
  {NoStop}%
\bibitem [{\citenamefont {Oliveira}(2019)}]{vol12}%
  \BibitemOpen
  \bibfield  {author} {\bibinfo {author} {\bibfnamefont {R.~R.~S.}\
  \bibnamefont {Oliveira}},\ }\href {\doibase 10.1140/epjc/s10052-019-7237-y}
  {\bibfield  {journal} {\bibinfo  {journal} {Eur. Phys. J. C}\ }\textbf
  {\bibinfo {volume} {79}},\ \bibinfo {pages} {725} (\bibinfo {year}
  {2019})}\BibitemShut {NoStop}%
\bibitem [{\citenamefont {Figueiredo}\ \emph {et~al.}(2017)\citenamefont
  {Figueiredo}, \citenamefont {Moraes}, \citenamefont {Fumeron},\ and\
  \citenamefont {Berche}}]{mmaterial1}%
  \BibitemOpen
  \bibfield  {author} {\bibinfo {author} {\bibfnamefont {D.}~\bibnamefont
  {Figueiredo}}, \bibinfo {author} {\bibfnamefont {F.}~\bibnamefont {Moraes}},
  \bibinfo {author} {\bibfnamefont {S.}~\bibnamefont {Fumeron}}, \ and\
  \bibinfo {author} {\bibfnamefont {B.}~\bibnamefont {Berche}},\ }\href@noop {}
  {\bibfield  {journal} {\bibinfo  {journal} {Phy. Rev. D}\ }\textbf {\bibinfo
  {volume} {96}},\ \bibinfo {pages} {105012} (\bibinfo {year}
  {2017})}\BibitemShut {NoStop}%
\bibitem [{\citenamefont {Nelson}(2002)}]{defects}%
  \BibitemOpen
  \bibfield  {author} {\bibinfo {author} {\bibfnamefont {D.~R.}\ \bibnamefont
  {Nelson}},\ }\href@noop {} {\emph {\bibinfo {title} {Defects and geometry in
  condensed matter physics}}}\ (\bibinfo  {publisher} {Cambridge University
  Press},\ \bibinfo {year} {2002})\BibitemShut {NoStop}%
\bibitem [{\citenamefont {Bezerra}\ \emph {et~al.}(2019)\citenamefont
  {Bezerra}, \citenamefont {Lobo}, \citenamefont {Mota},\ and\ \citenamefont
  {Muniz}}]{rainbow10}%
  \BibitemOpen
  \bibfield  {author} {\bibinfo {author} {\bibfnamefont {V.~B.}\ \bibnamefont
  {Bezerra}}, \bibinfo {author} {\bibfnamefont {I.~P.}\ \bibnamefont {Lobo}},
  \bibinfo {author} {\bibfnamefont {H.~F.}\ \bibnamefont {Mota}}, \ and\
  \bibinfo {author} {\bibfnamefont {C.~R.}\ \bibnamefont {Muniz}},\ }\href@noop
  {} {\bibfield  {journal} {\bibinfo  {journal} {Ann. Phys.}\ }\textbf
  {\bibinfo {volume} {401}},\ \bibinfo {pages} {162} (\bibinfo {year}
  {2019})}\BibitemShut {NoStop}%
\bibitem [{\citenamefont {Bakke}\ and\ \citenamefont {Mota}(2018)}]{rainbow9}%
  \BibitemOpen
  \bibfield  {author} {\bibinfo {author} {\bibfnamefont {K.}~\bibnamefont
  {Bakke}}\ and\ \bibinfo {author} {\bibfnamefont {H.}~\bibnamefont {Mota}},\
  }\href@noop {} {\bibfield  {journal} {\bibinfo  {journal} {The European
  Physical Journal Plus}\ }\textbf {\bibinfo {volume} {133}},\ \bibinfo {pages}
  {409} (\bibinfo {year} {2018})}\BibitemShut {NoStop}%
\bibitem [{\citenamefont {Kibble}(1976)}]{string15}%
  \BibitemOpen
  \bibfield  {author} {\bibinfo {author} {\bibfnamefont {T.~W.~B.}\
  \bibnamefont {Kibble}},\ }\href {\doibase 10.1088/0305-4470/9/8/029}
  {\bibfield  {journal} {\bibinfo  {journal} {J. Phys. A}\ }\textbf {\bibinfo
  {volume} {9}},\ \bibinfo {pages} {1387} (\bibinfo {year} {1976})}\BibitemShut
  {NoStop}%
\bibitem [{\citenamefont {Momeni}\ \emph {et~al.}(2017)\citenamefont {Momeni},
  \citenamefont {Upadhyay}, \citenamefont {Myrzakulov},\ and\ \citenamefont
  {Myrzakulov}}]{rainbow12}%
  \BibitemOpen
  \bibfield  {author} {\bibinfo {author} {\bibfnamefont {D.}~\bibnamefont
  {Momeni}}, \bibinfo {author} {\bibfnamefont {S.}~\bibnamefont {Upadhyay}},
  \bibinfo {author} {\bibfnamefont {Y.}~\bibnamefont {Myrzakulov}}, \ and\
  \bibinfo {author} {\bibfnamefont {R.}~\bibnamefont {Myrzakulov}},\
  }\href@noop {} {\bibfield  {journal} {\bibinfo  {journal} {Astrophys. Space
  Sci.}\ }\textbf {\bibinfo {volume} {362}},\ \bibinfo {pages} {148} (\bibinfo
  {year} {2017})}\BibitemShut {NoStop}%
\bibitem [{\citenamefont {Abramowitz}\ and\ \citenamefont
  {Stegun}(1964)}]{abramo}%
  \BibitemOpen
  \bibfield  {author} {\bibinfo {author} {\bibfnamefont {M.}~\bibnamefont
  {Abramowitz}}\ and\ \bibinfo {author} {\bibfnamefont {I.}~\bibnamefont
  {Stegun}},\ }\href {https://books.google.com.br/books?id=MtU8uP7XMvoC} {\emph
  {\bibinfo {title} {Handbook of Mathematical Functions: With Formulas, Graphs,
  and Mathematical Tables}}},\ Applied mathematics series\ (\bibinfo
  {publisher} {Dover Publications},\ \bibinfo {year} {1964})\BibitemShut
  {NoStop}%
\bibitem [{\citenamefont {Magueijo}\ and\ \citenamefont
  {Smolin}(2002)}]{rainbow2}%
  \BibitemOpen
  \bibfield  {author} {\bibinfo {author} {\bibfnamefont {J.}~\bibnamefont
  {Magueijo}}\ and\ \bibinfo {author} {\bibfnamefont {L.}~\bibnamefont
  {Smolin}},\ }\href@noop {} {\bibfield  {journal} {\bibinfo  {journal} {Phys.
  Rev. Lett.}\ }\textbf {\bibinfo {volume} {88}},\ \bibinfo {pages} {190403}
  (\bibinfo {year} {2002})}\BibitemShut {NoStop}%
\bibitem [{\citenamefont {Ali}(2014)}]{rainbow27}%
  \BibitemOpen
  \bibfield  {author} {\bibinfo {author} {\bibfnamefont {A.~F.}\ \bibnamefont
  {Ali}},\ }\href {\doibase 10.1103/PhysRevD.89.104040} {\bibfield  {journal}
  {\bibinfo  {journal} {Phys. Rev. D}\ }\textbf {\bibinfo {volume} {89}},\
  \bibinfo {pages} {104040} (\bibinfo {year} {2014})},\ \Eprint
  {http://arxiv.org/abs/1402.5320} {arXiv:1402.5320 [hep-th]} \BibitemShut
  {NoStop}%
\end{thebibliography}%

$\bigskip $

\end{document}